\documentclass[oneside,10pt]{article}

\usepackage{times,url,mathrsfs}
\usepackage{amsmath}
\usepackage{amsfonts}
\usepackage{graphicx}
\usepackage{subfigure}
\newtheorem{theorem}{Theorem}
\newtheorem{lemma}{Lemma}

\begin{document}

\title{On the Sample Complexity of Compressed Counting}

\author{ Ping Li \\
         Department of Statistical Science\\
         Faculty of Computing and Information Science\\
       Cornell University\\
         Ithaca, NY 14853\\
       pingli@cornell.edu
  }
\date{July 6, 2009}
\maketitle

\begin{abstract}

Compressed\footnote{Extended abstract, submitted on July 6, 2009.}
 Counting (CC)\cite{Proc:Li_SODA09}, based on {\em maximally skewed stable random projections}, was recently proposed for  estimating the $\alpha$th frequency moments of data streams. When $\Delta=|1-\alpha|\rightarrow 0$, \cite{Proc:Li_SODA09} provided an algorithm based on the {\em geometric mean} estimator and proved that the sample complexity was essentially $O\left(1/\epsilon\right)$, which was a  large improvement compared to the previously known $O\left(1/\epsilon^2\right)$ bound. The case $\Delta=|1-\alpha|\rightarrow0$ is extremely useful for estimating Shannon entropy of data streams.

In this study, we provide a very simple algorithm based on the {\em sample minimum} estimator and prove that, when $\Delta = 1-\alpha\rightarrow 0+$,  it suffices to let the sample size $k$ be
\begin{align}\notag
k \geq \frac{\log\frac{1}{\delta}}{\log\frac{1}{\Delta} -\log\left(
\frac{1}{2}+\frac{1}{2\log(1+\epsilon)}+\frac{1}{2\Delta\log\Delta+2\log(1+\epsilon)}+O\left(\Delta\right)
\right)}
\end{align}
so that, with probability at least $1-\delta$, the estimated $\alpha$th frequency moments will be within a $1+\epsilon$ factor of the truth. For example, when $\epsilon=10^{-3}$, $\delta=10^{-10}$, and $\Delta=10^{-5}$, the required sample size is merely $k\geq 5.1$.

\end{abstract}


\section{Introduction}

The problem of ``scaling up for high dimensional data and high speed data streams'' is among the  ``ten challenging problems in data mining research''\cite{Article:ICDM10}. This paper is devoted to estimating entropy of data streams. Mining data streams\cite{Book:Henzinger_99,Proc:Babcock_PODS02,Proc:Aggarwal_KDD04,Article:Muthukrishnan_05} in (e.g.,) 100 TB scale  databases has become an important area of research, e.g., \cite{Proc:Domeniconi_ICDM01,Proc:Aggarwal_KDD04}, as network data can easily reach that scale\cite{Article:ICDM10}. Search engines are a typical source of data streams\cite{Proc:Babcock_PODS02}.

Consider the  {\em Turnstile}  stream model\cite{Article:Muthukrishnan_05}. The input  stream $a_t = (i_t, I_t)$, $i_t\in [1,\ D]$ arriving sequentially describes the underlying signal $A$, meaning
\begin{align}\label{eqn_Turnstile}
A_t[i_t] = A_{t-1}[i_t] + I_t,
\end{align} where the increment $I_t$ can be either positive (insertion) or negative (deletion).  Restricting $A_t[i]\geq 0$ results in the {\em strict-Turnstile} model, which suffices for describing almost all natural phenomena.
This study focuses on the {\em strict-Turnstile} model and  studies efficient algorithms for estimating the {\em $\alpha$th frequency moments} of data streams
\begin{align}\label{eqn_moment}
F_{(\alpha)} = \sum_{i=1}^D A_t[i]^\alpha.
\end{align}
We are particularly interested in the case of $\alpha\rightarrow 1$, which is very important for estimating {\em Shannon entropy}.

\subsection{Entropy}

A very useful (e.g.,  in Web and networks\cite{Proc:Feinstein_DARPA03,Proc:Lall_SIGMETRICS06,Proc:Zhao_IMC07,Proc:Mei_WSDM08} and neural comptutations\cite{Article:Paninski_NC03}) summary statistic is  the {\em Shannon entropy}
{\small\begin{align}\label{eqn_Shannon}
H = -\sum_{i=1}^D\frac{A_t[i]}{F_{(1)}}\log \frac{A_t[i]}{F_{(1)}}.
\end{align}}
Various generalizations of the Shannon entropy have been proposed. The R\'enyi entropy\cite{Proc:Renyi_61}, denoted by $H_\alpha$, and the Tsallis entropy\cite{Article:Havrda_67,Article:Tsallis_88}, denoted by $T_\alpha$,   are respectively defined as
{\small\begin{align}\label{eqn_Renyi}
&H_\alpha =\frac{1}{1-\alpha} \log \frac{\sum_{i=1}^D A_t[i]^\alpha}{\left(\sum_{i=1}^D A_t[i]\right)^\alpha}, \hspace{0.5in}T_\alpha = \frac{1}{\alpha -1} \left( 1 - \frac{F_{(\alpha)}}{F_{(1)}^\alpha}\right).
\end{align}}

As $\alpha\rightarrow 1$, both R\'enyi entropy and Tsallis entropy converge to Shannon entropy:
$\lim_{\alpha\rightarrow 1} H_\alpha  = \lim_{\alpha\rightarrow 1}T_\alpha = H$.  Thus, both R\'enyi entropy and Tsallis entropy can be computed from the $\alpha$th frequency moment; and one can approximate Shannon entropy from either $H_\alpha$ or $T_\alpha$ by letting $\alpha\approx 1$. Several studies\cite{Proc:Zhao_IMC07,Proc:Harvey_ITW08,Proc:Harvey_FOCS08})   used this idea to approximate  Shannon entropy, all of which relied on efficient algorithms for estimating the $\alpha$th estimating frequency moments (\ref{eqn_moment}) near $\alpha =1$. In fact, one can numerically verify that the $\alpha$ values proposed in \cite{Proc:Harvey_ITW08,Proc:Harvey_FOCS08} are extremely close to 1, e.g., $\Delta = |1-\alpha|\leq 10^{-4}$.

Therefore, efficient algorithms for estimating $F_{(\alpha)}$ near $\alpha=1$ is critical for estimating Shannon entropy.

\subsection{Sample Applications of Shannon Entropy}

\subsubsection{Real-Time Network Anomaly Detection}

Network traffic is a typical example of high-rate data streams. An effective and reliable measurement of network traffic in real-time is crucial for anomaly detection and network diagnosis; and one such measurement metric is Shannon entropy\cite{Proc:Feinstein_DARPA03,Proc:Lakhina_SIGCOMM05,Proc:Xu_SIGCOMM05,Proc:Brauckhoff_IMC06,Proc:Lall_SIGMETRICS06,Proc:Zhao_IMC07}. The {\em Turnstile} data stream model (\ref{eqn_Turnstile}) is naturally suitable for describing network traffic, especially when the goal is to characterize the statistical distribution of the traffic. In its empirical form, a statistical distribution is described by histograms, $A_t[i]$, $i=1$ to $D$. It is possible that $D=2^{64}$ (IPV6) if one is interested in measuring the traffic streams of unique source or destination.

The Distributed Denial of Service (\textbf{DDoS}) attack is a representative example of network anomalies. A DDoS attack attempts to make computers unavailable to intended users, either by forcing users to reset the computers or by exhausting the resources of  service-hosting sites. For example, hackers may maliciously saturate the victim machines by sending many external communication requests. DDoS attacks typically target sites such as banks, credit card payment gateways, or military sites.

A DDoS attack changes the statistical distribution of network traffic. Therefore, a common practice to detect an attack is to monitor the network traffic using certain summary statics. Since  Shannon entropy is a well-suited for characterizing a distribution, a popular detection method is to measure the time-history of entropy and alarm anomalies when the entropy becomes abnormal\cite{Proc:Feinstein_DARPA03,Proc:Lall_SIGMETRICS06}.

Entropy measurements do not have to be ``perfect'' for detecting attacks. It is however crucial that the  algorithm should be computationally efficient at  low memory cost, because the traffic data generated by large high-speed networks are enormous and transient (e.g., 1 Gbits/second). Algorithms should be real-time and one-pass, as the traffic data will not be stored\cite{Proc:Babcock_PODS02}. Many algorithms have been proposed for ``sampling'' the traffic data and estimating entropy over data streams\cite{Proc:Lall_SIGMETRICS06,Proc:Zhao_IMC07,Proc:Bhuvanagiri_ESA06,Proc:Guha_SODA06,Article:Chakrabarti_06,Proc:Chakrabarti_SODA07,Proc:Harvey_ITW08,Proc:Harvey_FOCS08},

\subsubsection{Entropy of Query Logs in Web Search}
The recent work\cite{Proc:Mei_WSDM08} was devoted to estimating the Shannon entropy of MSN search logs, to help answer some basic problems in Web search, such as,  {\em how big is the web?}

The search logs can be viewed as data streams, and \cite{Proc:Mei_WSDM08}  analyzed several ``snapshots'' of a sample of MSN search logs.  The sample used in \cite{Proc:Mei_WSDM08} contained 10 million $<$Query, URL,IP$>$ triples; each triple corresponded to a click from a particular IP address on a particular URL for a particular query.  \cite{Proc:Mei_WSDM08} drew their important conclusions on this (hopefully) representative sample. Alternatively, one could apply data  stream algorithms such as CC on the whole history of MSN (or other search engines).

\vspace{-0.05in}
\subsubsection{Entropy in Neural Computations}

A workshop in NIPS'03 was denoted to entropy estimation, owing to the wide-spread use of Shannon entropy in Neural Computations\cite{Article:Paninski_NC03}. (\url{http://www.menem.com/~ilya/pages/NIPS03}) For example, one application of entropy is to study the underlying structure of  spike trains.

\subsection{Previous Algorithms for Estimating Frequency Moments}

The problem of approximating $F_{(\alpha)}$ has been very heavily studied in theoretical computer science and databases, since the pioneering work of \cite{Proc:Alon_STOC96}, which studied $\alpha = 0$,  2, and $\alpha>2$. \cite{Proc:Feigenbaum_FOCS99,Article:Indyk_JACM06,Proc:Li_SODA08} provided improved algorithms for $0<\alpha\leq 2$. \cite{Proc:Indyk_STOC05} provided algorithms for $\alpha >2$ to  achieve the lower bounds proved by
 \cite{Proc:Saks_STOC02,Proc:Kumar_FOCS02,Proc:Woodruff_SODA04}. \cite{Proc:Ganguly_RANDOM07} suggested using even more space to trade for some speedup in the processing time.

Note that the first moment (i.e., the sum), $F_{(1)}$, can be computed easily with a simple counter\cite{Article:Morris_CACM78,Article:Flajolet_BIT85,Proc:Alon_STOC96}. This important property was recently somewhat captured by the method of {\em Compressed Counting (CC)}\cite{Proc:Li_SODA09}, which was based on the {\em maximally-skewed stable random projections}.
\cite{Proc:Li_SODA09} proved that, in the neighborhood of $\alpha=1$, the sample complexity is essentially $O\left(1/\epsilon\right)$, which was a large improvement over the well-known $O\left(1/\epsilon^2\right)$ bound\cite{Proc:Woodruff_SODA04,Article:Indyk_JACM06,Proc:Li_SODA08}. This means the required sample size using CC should be $O\left(1/\epsilon\right)$ in order to ensure that the estimated $\alpha$th frequency moment will be within a $1\pm\epsilon$ factor of the truth, with high probability. \\

The sample complexity bound of $O\left(1/\epsilon\right)$  for CC is  unsatisfactory, not just for theoretical reasons. From a practical point of view, $1/\epsilon$ can be too large to be practical, especially for entropy estimation. For example, one can numerically verify that the required $\epsilon$ values in \cite{Proc:Harvey_ITW08,Proc:Harvey_FOCS08} for entropy estimation are very small. Very recently, without providing any theoretical complexity bounds, \cite{Proc:Li_UAI09} proposed an empirically improved (and quite sophisticated) algorithm for CC. Because the algorithm in \cite{Proc:Li_UAI09} is quite complex, its theoretical analysis was difficult.

This study proposes a very simple algorithm, which also allows us to analyze its sample complexity. The complexity is essentially $O\left(\frac{1}{\log\left(1/\Delta\right)-\log(1/\epsilon)}\right)$, when $\Delta=1-\alpha\rightarrow 0$.

\section{The Proposed Algorithm and Main Theoretical Results}

We consider the {\em strict-Turnstile} model (\ref{eqn_Turnstile}). Conceptually, we  multiply the data stream vector $A_t\in\mathbb{R}^{1\times D}$ by a random projection matrix $\mathbf{R}\in\mathbb{R}^{D\times k}$. The resultant vector $X = A_t \times \mathbf{R}\in\mathbb{R}^{k\times 1}$ is only of length $k$.  More specifically, the entries of the projected vector $X$ are
\begin{align}\notag
x_j =\left[A_t\times\mathbf{R}\right]_j= \sum_{i=1}^D r_{ij} A_t[i], \ \ j = 1, 2, ..., k
\end{align}

$r_{ij}$'s are random variables generated by
\begin{align}\label{eqn_r_ij}
r_{ij} =  \frac{  \sin\left(\alpha v_{ij} \right)}{\left[\sin v_{ij}
\right]^{1/\alpha}} \left[\frac{\sin\left( v_{ij}\Delta\right)}{w_{ij}}
\right]^{\frac{\Delta}{\alpha}}, \ \ \ \ \Delta = 1-\alpha>0,
\end{align}
where $v_{ij} \sim uniform(0,\pi)$ (i.i.d.) and $w_{ij}\sim \exp(1)$ (i.i.d.), an exponential distribution with mean 1.

Of course, in data stream computations, the matrix $\mathbf{R}$ is never fully materialized. The standard procedure in data stream computations is to generate entries of $\mathbf{R}$ on-demand\cite{Article:Indyk_JACM06}. In other words, whenever an stream element $a_t = (i_t, I_t)$ arrives, one updates entries of $X$ as
\begin{align}\notag
x_j \leftarrow x_j + I_t r_{i_tj},  \ \ \ j = 1, 2, ..., k.
\end{align}

The proposed algorithm is to take the {\em sample minimum}:
\begin{align}\label{eqn_F_min}
\hat{F}_{(\alpha),\min} = \left[\min \left\{ x_j, \ j = 1, 2, ..., k\right\}\right]^\alpha.
\end{align}
While this estimator is extremely simple, it has  nice theoretical properties.

\begin{theorem}\label{thm_right_bound}

As $\Delta=1-\alpha \rightarrow 0+$, for any fixed $\epsilon >0$,
\begin{align}\label{eqn_right_bound}
&\mathbf{Pr}\left(\hat{F}_{(\alpha),\min} \geq (1+\epsilon)F_{(\alpha)}\right)\leq
\exp\left(k \log\frac{1}{2}\left[
\Delta+\frac{\Delta}{\log(1+\epsilon)}+\frac{\Delta}{\Delta\log\Delta+\log(1+\epsilon)}+O\left(\Delta^2\right)
 \right]\right)
\end{align}

Therefore, it suffices to let the sample size
\begin{align}\label{eqn_complexity}
k \geq \frac{\log\frac{1}{\delta}}{\log\frac{1}{\Delta} -\log\left(
\frac{1}{2}+\frac{1}{2\log(1+\epsilon)}+\frac{1}{2\Delta\log\Delta+2\log(1+\epsilon)}+O\left(\Delta\right)
\right)}
\end{align}
so that with probability at least $1-\delta$, $\hat{F}_{(\alpha),\min}$ is within a $1+\epsilon$ factor of $F_{(\alpha)}$.

\end{theorem}

The proof is deferred to Section \ref{sec_proof_thm_right_bound}, which will also demonstrate that the right tail bound (\ref{eqn_right_bound}) can be slightly improved by essentially removing the $\Delta\log\Delta$ term in (\ref{eqn_right_bound}).

\begin{figure}[h]
\begin{center}\mbox{
\includegraphics[width=2.5in]{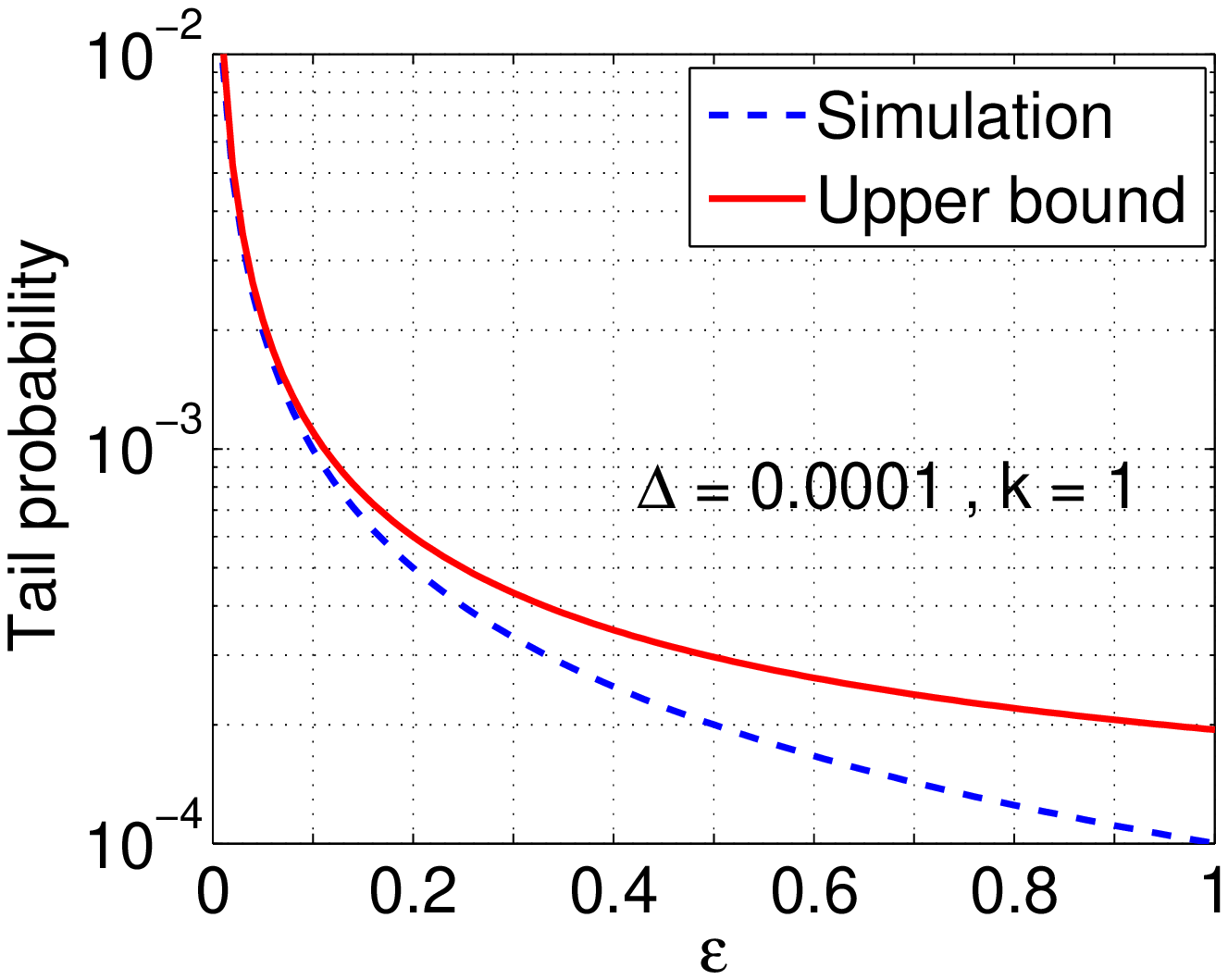}
\includegraphics[width=2.5in]{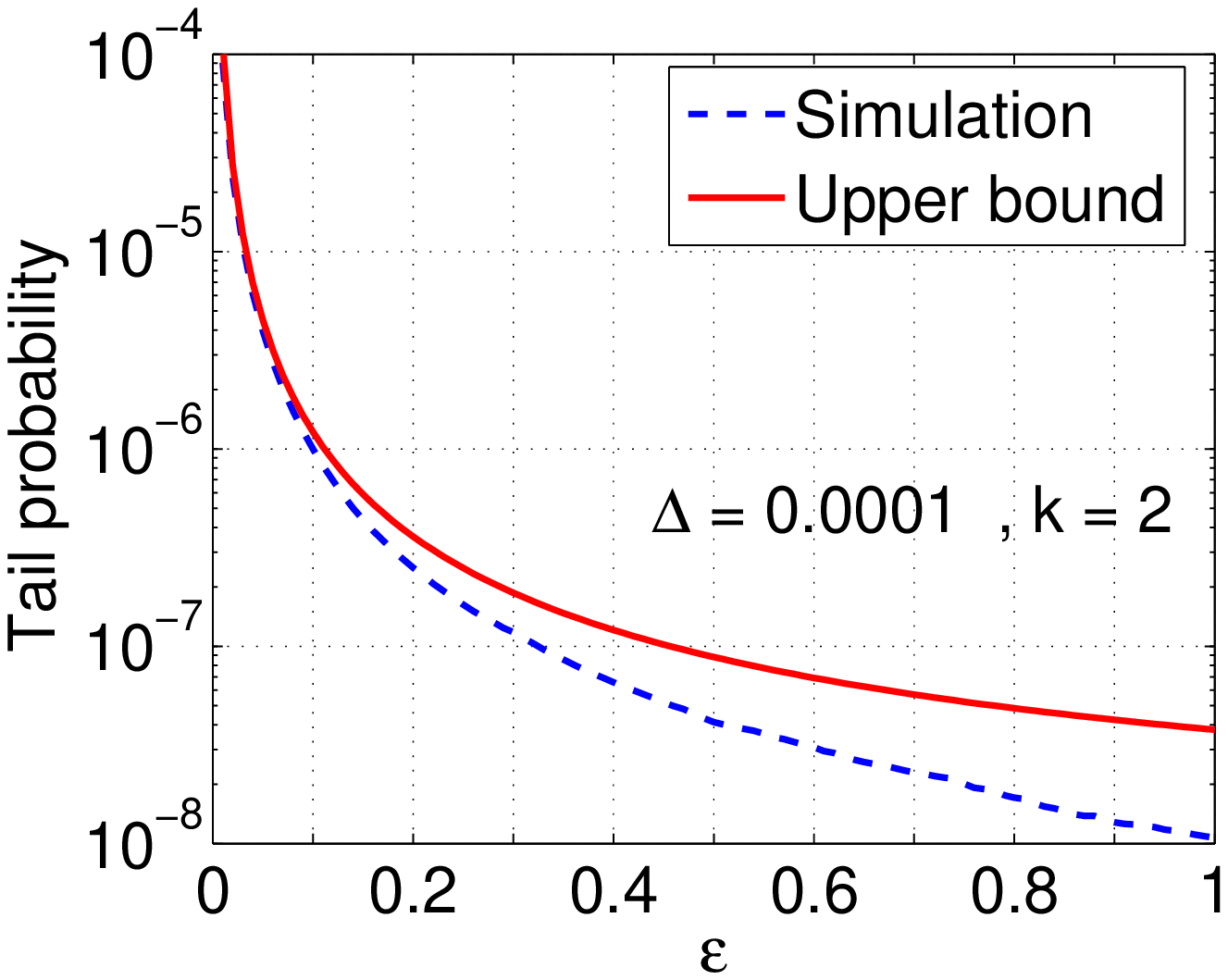}}
\mbox{
\includegraphics[width=2.5in]{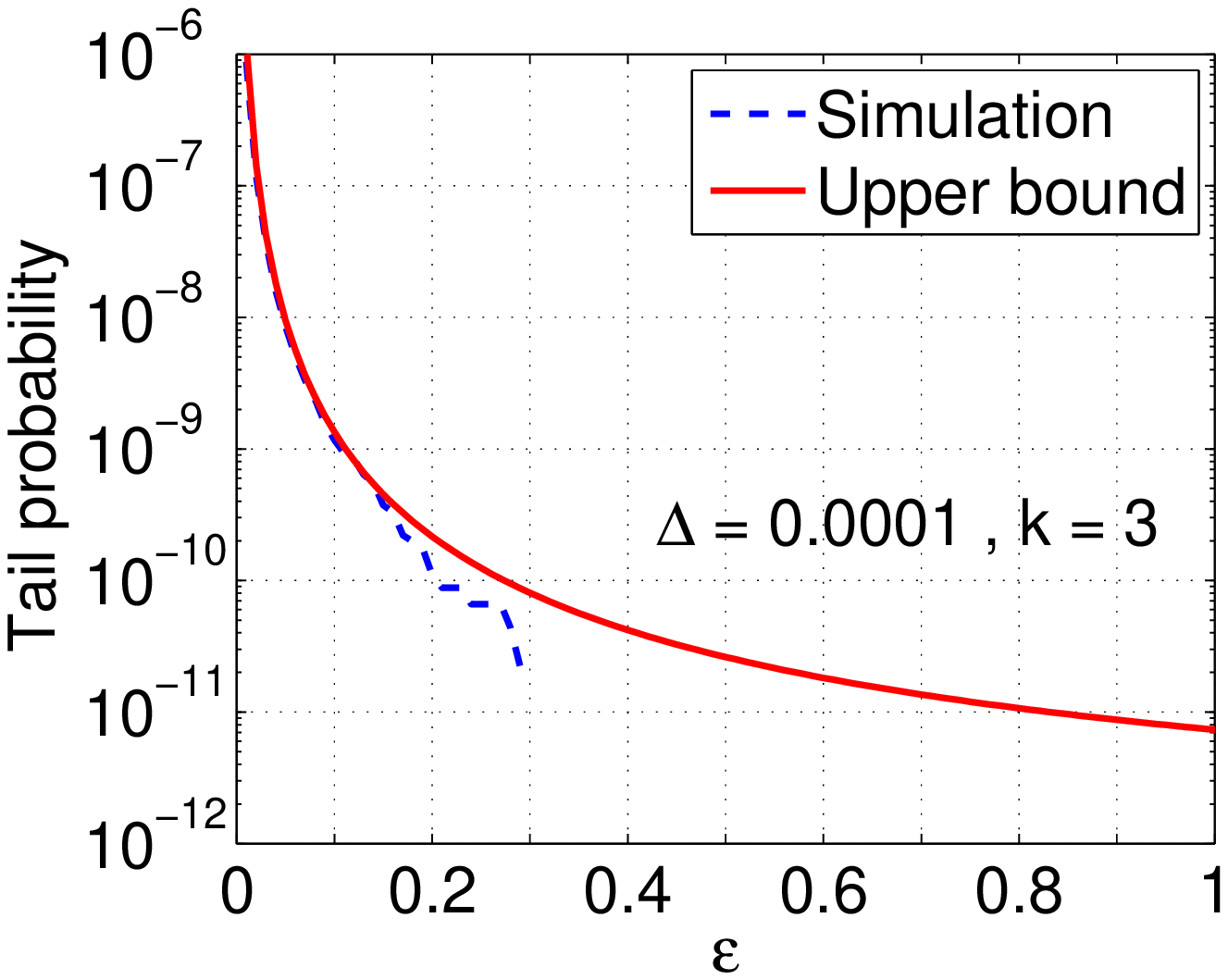}
\includegraphics[width=2.5in]{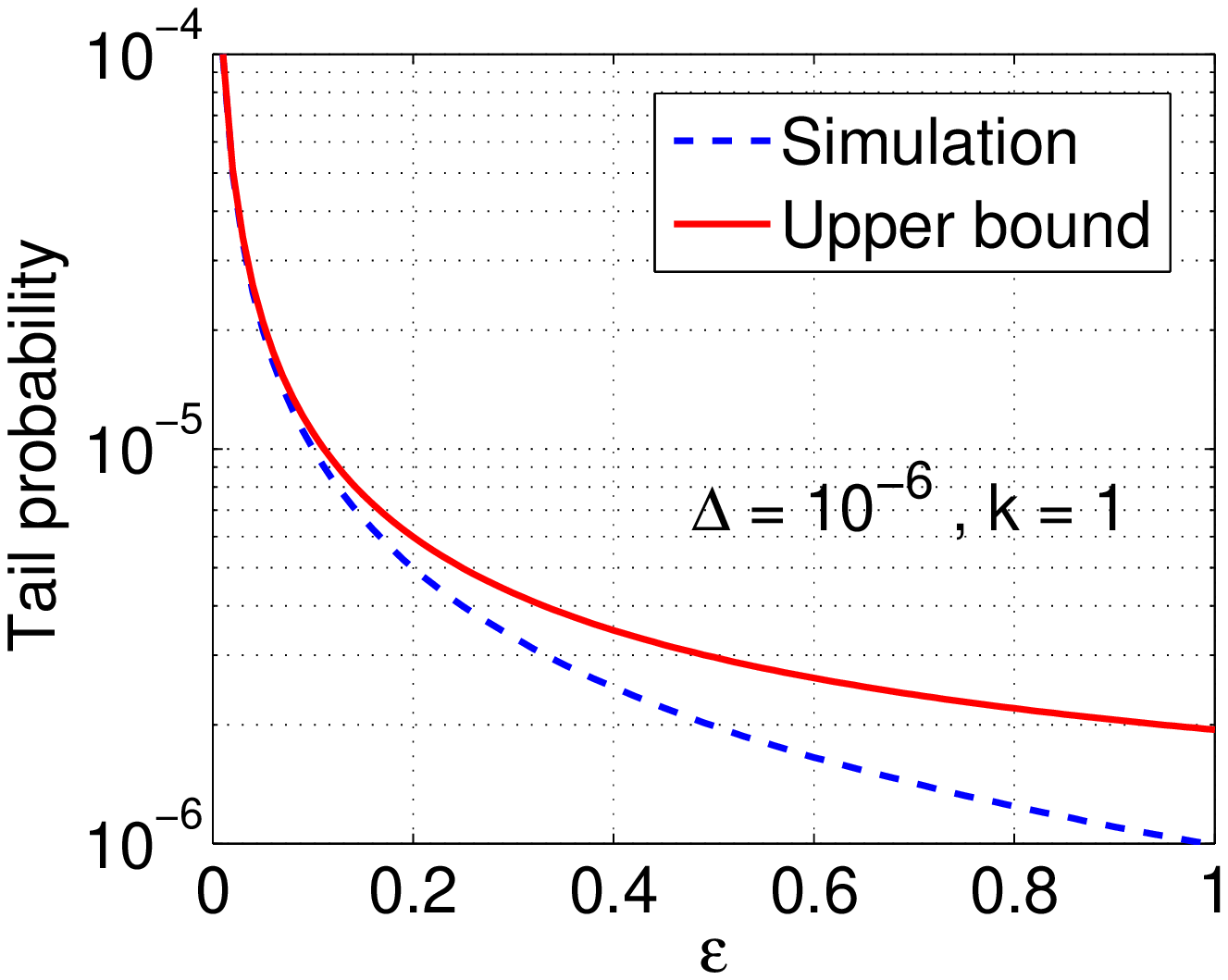}
}
\end{center}
\vspace{-0.in}
\caption{Right tail bound (\ref{eqn_right_bound}) for selected $\Delta$ and $k$, together with the simulated tail probabilities. }\label{fig_right_bound}
\end{figure}

To help verify the results in Theorem \ref{thm_right_bound}, Figure \ref{fig_right_bound} plots the right tail bounds (\ref{eqn_right_bound}) for $\Delta = 10^{-4}$ ($k =1, 2, 3$) and $\Delta=10^{-6}$ ($k=1$ only), together with the simulated tail probabilities. We can see that the tail probabilities decrease very rapidly. In fact, it is even difficult to simulate the tail probabilities if $k>3$ or $\Delta<10^{-6}$.

Theorem \ref{thm_right_bound} indicates that required sample size $k$ can be very small. For example, if we let $\epsilon=10^{-3}$, $\delta=10^{-10}$, and $\Delta=10^{-5}$, then according to (\ref{eqn_complexity}), the required sample size is merely $k\geq 5.1$

Note that Theorem \ref{thm_right_bound} is just for  the sample complexity. To obtain the space complexity, we must consider an multiplicative factor of $\log \sum_{s=1}^t |I_s|$. In addition, we must store $r_{ij}$ with a sufficient accuracy. In Section \ref{sec_preproofs}, Lemma \ref{lem_Z_order} shows that $\log r_{ij} = O\left(|\Delta\log\Delta|\right)$, which can be represented using $O\left(\log1/\Delta\right)$ bits. Therefore, the required storage space would be the sample complexity (\ref{eqn_complexity}) multiplied by a factor of $O\left(\log \sum_{s=1}^t |I_s|\right)$ +$O\left(\log1/\Delta\right)$. \\

\newpage

Theorem \ref{thm_left_bound} presents the left tail bound.
\begin{theorem}\label{thm_left_bound}
For any $0<\epsilon<1$,  $\alpha<1$, and $\Delta = 1-\alpha$,
\begin{align}\label{eqn_left_bound}
&\mathbf{Pr}\left(\hat{F}_{(\alpha),\min} \leq (1-\epsilon)F_{(\alpha)}\right)
\leq k\exp\left(-\frac{\Delta\alpha^{1/\Delta-1}}{(1-\epsilon)^{1/\Delta}}\right).
\end{align}
\end{theorem}
The proof is deferred to Section \ref{sec_proof_thm_left_bound}.

The left bound (\ref{eqn_left_bound}) approaches zero extremely fast. For example, when $\Delta=10^{-6}$ and $\epsilon=10^{-4}$, $\frac{\Delta\alpha^{1/\Delta-1}}{(1-\epsilon)^{1/\Delta}}\approx 10^{37}$; and hence $k$ does not really matter for the left bound. In a  sense, the left bound will be used merely for the sanity check and one can determine the sample size mainly from the right bound in Theorem \ref{thm_right_bound}.

\section{Preparation for the Proofs of the Main Results}\label{sec_preproofs}

We start with reviewing maximally-skewed stable distributions, because our formulation (\ref{eqn_r_ij}) somewhat differs from the standard formulation.

\subsection{Maximally-Skewed Stable Distribution}

The standard procedure for sampling from skewed stable distributions is based on the Chambers-Mallows-Stuck method\cite{Article:Chambers_JASA76}. To generate a sample from $S(\alpha,\beta=1,1)$, i.e., $\alpha$-stable, maximally-skewed ($\beta=1$), with unit scale,  one first generates an exponential random variable with mean 1, $W \sim \exp(1)$,  and a uniform random variable $U \sim uniform \left(-\frac{\pi}{2}, \frac{\pi}{2}\right)$, then,
\begin{align}\label{eqn_sampling_skewed}
Z^\prime &= \frac{\sin\left(\alpha(U+\rho)\right)}{\left[\cos U \cos\left(\rho \alpha\right)
\right]^{1/\alpha}} \left[\frac{\cos\left( U - \alpha(U + \rho)\right)}{W}
\right]^{\frac{1-\alpha}{\alpha}} \sim S(\alpha,\beta=1,1),
\end{align}
where $\rho = \frac{\pi}{2}$ when $\alpha<1$ and $\rho = \frac{\pi}{2}\frac{2-\alpha}{\alpha}$ when $\alpha>1$. \\

For convenience, we will use
\begin{align}\notag
Z = Z^\prime\cos^{1/\alpha}\left(\rho\alpha\right) \sim S\left(\alpha,\beta=1,\cos\left(\rho\alpha\right)\right).
\end{align}

In this study, we will only consider\ \  $\alpha = 1-\Delta<1$, i.e, $\rho=\frac{\pi}{2}$. After simplification, we obtain
\begin{align}
Z &= \frac{  \sin\left(\alpha V \right)}{\left[\sin V
\right]^{1/\alpha}} \left[\frac{\sin\left( V\Delta\right)}{W}
\right]^{\frac{\Delta}{\alpha}},
\end{align}
where $V = \frac{\pi}{2} + U\sim uniform(0,\pi)$. This explains (\ref{eqn_r_ij}).\\

Lemma \ref{lem_Z_order} shows $\log Z = O\left(|\Delta\log\Delta|\right)$, which can be accurately represented using $O\left(\log 1/\Delta\right)$ bits. The proof is omitted since it is straightforward.

\begin{lemma}\label{lem_Z_order}
For any given $V\neq 0$, and $W\neq 0$, as $\Delta\rightarrow 0$,
\begin{align}\notag
Z = 1+O\left(|\Delta\log\Delta|\right), \ \ \ \text{i.e.,} \ \ \log Z = O\left(|\Delta\log\Delta|\right).
\end{align}
\end{lemma}

\subsection{Random Projections and the Sample Minimum Estimator}

Let $X = A_t\times \mathbf{R}$, where entries are $\mathbf{R}$ are i.i.d. samples of $S\left(\alpha,\beta=1,\cos\left(\frac{\pi}{2}\alpha\right)\right)$. Then by properties of stable distributions, entries of $X$ are
\begin{align}\notag
x_j = \left[A_t\times \mathbf{R}\right]_j = \sum_{i=1}^D r_{i,j}A_t[i] \sim S \left(\alpha,\beta=1, \cos\left(\frac{\pi}{2}\alpha\right)F_{(\alpha)}\right),
\end{align}
where $F_{(\alpha)}=\sum_{i=1}^D A_t[i]^\alpha$ as defined in (\ref{eqn_moment}).

The proposed estimator of $F_{(\alpha)}$ is based on the {\em sample minimum}:
\begin{align}\notag
\hat{F}_{(\alpha),\min} = \left[\min\left\{x_j, j = 1, 2, ..., k\right\}\right]^\alpha
\end{align}

\subsection{Density Function}

\begin{lemma}\label{lem_CDF}
Suppose a random variable $Z \sim S\left(\alpha<1,\beta=1,\cos\left(\frac{\pi}{2}\alpha\right)\right)$,
then the cumulative density function is
\begin{align}\notag
&F_Z(t) =  \mathbf{Pr}\left(Z\leq t\right) =\frac{1}{\pi}\int_0^\pi  \exp\left(- \frac{  \left[\sin\left(\alpha \theta \right)\right]^{\alpha/\Delta} }{t^{\alpha/\Delta} \left[\sin \theta \right]^{1/\Delta}} \sin\left( \theta\Delta\right)\right) d\theta, \hspace{0.5in} (\Delta = 1-\alpha).
\end{align}

\textbf{Proof:}
\begin{align}\notag
&\mathbf{Pr}\left(Z\geq t\right) =\mathbf{Pr}\left( \frac{  \sin\left(\alpha V \right)}{\left[\sin V
\right]^{1/\alpha}} \left[\frac{\sin\left( V\Delta\right)}{W}
\right]^{\frac{\Delta}{\alpha}}  \geq t \right)\\\notag
=&\mathbf{Pr}\left(  W \leq \frac{  \left[\sin\left(\alpha V \right)\right]^{\alpha/\Delta} }{t^{\alpha/\Delta} \left[\sin V
\right]^{1/\Delta}} \sin\left( V\Delta\right)\right)\\\notag
=&\text{E}\left(\mathbf{Pr}\left( \left. W \leq \frac{  \left[\sin\left(\alpha V \right)\right]^{\alpha/\Delta} }{t^{\alpha/\Delta} \left[\sin V
\right]^{1/\Delta}} \sin\left( V\Delta\right)\right|V\right)\right)\\\notag
=&1-\text{E}\left(\exp\left(- \frac{  \left[\sin\left(\alpha V \right)\right]^{\alpha/\Delta} }{t^{\alpha/\Delta} \left[\sin V \right]^{1/\Delta}} \sin\left( V\Delta\right)\right)\right)\\\notag
=&1-\frac{1}{\pi}\int_0^\pi  \exp\left(- \frac{  \left[\sin\left(\alpha \theta \right)\right]^{\alpha/\Delta} }{t^{\alpha/\Delta} \left[\sin \theta \right]^{1/\Delta}} \sin\left( \theta\Delta\right)\right) d\theta\Box
\end{align}

\end{lemma}

For $\theta\in(0,\pi)$, let
\begin{align}\notag
g(\theta;\Delta) = \frac{  \left[\sin\left(\alpha \theta \right)\right]^{\alpha/\Delta} }{\left[\sin \theta \right]^{1/\Delta}} \sin\left( \theta\Delta\right),
\end{align}

Lemma \ref{lem_g} includes some properties of $g(\theta;\Delta)$, which will be useful for proving our main results in Theorem \ref{thm_right_bound} and Theorem \ref{thm_left_bound}.
\begin{lemma}\label{lem_g}
Assume $\Delta = 1-\alpha<0.5$, then
$g(\theta;\Delta)$ is monotonically increasing in $(0,\pi)$, with
\begin{align}\notag
\lim_{\theta\rightarrow0+}g(\theta;\Delta) = \Delta\alpha^{1/\Delta-1}.
\end{align}
Moreover, $g(\theta;\Delta)$ is a convex function of $\theta$.
\end{lemma}

\section{Proofs of Theorem \ref{thm_right_bound} and Theorem \ref{thm_left_bound}}

We first prove the left bound in Theorem \ref{thm_left_bound}.

\subsection{Proof of Theorem \ref{thm_left_bound}}\label{sec_proof_thm_left_bound}

Recall the {\em sample minimum} estimator is
\begin{align}\notag
\hat{F}_{(\alpha),\min} = \left[\min\left\{x_j, j = 1, 2, ..., k\right\}\right]^\alpha, \hspace{0.2in} x_j \sim S\left(\alpha<1, \beta=1, \cos\left(\frac{\pi}{2}\alpha\right)F_{(\alpha)}\right).
\end{align}

Using the density function provided in Lemma \ref{lem_CDF} and properties of  $g(\theta;\Delta) = \frac{  \left[\sin\left(\alpha \theta \right)\right]^{\alpha/\Delta} }{\left[\sin \theta \right]^{1/\Delta}} \sin\left( \theta\Delta\right)$ proved in Lemma \ref{lem_g} , we obtain
\begin{align}\notag
&\mathbf{Pr}\left(\hat{F}_{(\alpha),\min} \leq (1-\epsilon)F_{(\alpha)}\right)\\\notag
\leq& k\times \mathbf{Pr}\left(x_1^\alpha/F_{(\alpha)} \leq (1-\epsilon)\right)\\\notag
=&k\frac{1}{\pi}\int_0^\pi \exp\left(- \frac{  \left[\sin\left(\alpha \theta \right)\right]^{\alpha/\Delta} }{(1-\epsilon)^{1/\Delta} \left[\sin \theta \right]^{1/\Delta}} \sin\left( \theta\Delta\right)\right) d\theta\\\notag
\leq&k\frac{1}{\pi}\int_0^\pi \exp\left(- \frac{\lim_{\theta\rightarrow0+}g(\theta,\Delta)  }{(1-\epsilon)^{1/\Delta}}\right)  d\theta\\\notag
=&k\frac{1}{\pi}\int_0^\pi \exp\left(- \frac{\Delta\alpha^{1/\Delta-1}  }{(1-\epsilon)^{1/\Delta}}\right)  d\theta\\\notag
=& k\exp\left(-\frac{\Delta\alpha^{1/\Delta-1}}{(1-\epsilon)^{1/\Delta}}\right).
\end{align}

\subsection{Proof of Theorem \ref{thm_right_bound}}\label{sec_proof_thm_right_bound}

Using the density function provided in Lemma \ref{lem_CDF}, we can obtain
\begin{align}\notag
&\mathbf{Pr}\left(\hat{F}_{(\alpha),\min} \geq (1+\epsilon)F_{(\alpha)}\right)\\\notag=& \mathbf{Pr}\left(\hat{F}_{(\alpha),\min}/F_{(\alpha)} \geq (1+\epsilon)\right)\\\notag
=&\prod_{j=1}^k\mathbf{Pr}\left(x_j/F_{(\alpha)}^{1/\alpha} \geq (1+\epsilon)^{1/\alpha}\right)\\\notag
=&\left[1-\frac{1}{\pi}\int_0^\pi \exp\left(- \frac{  \left[\sin\left(\alpha \theta \right)\right]^{\alpha/\Delta} }{(1+\epsilon)^{1/\Delta} \left[\sin \theta \right]^{1/\Delta}} \sin\left( \theta\Delta\right)\right) d\theta
\right]^k\\\notag
=&\exp\left(k \log \left[1-\frac{1}{\pi}\int_0^\pi \exp\left(- \frac{ g(\theta;\Delta) }{(1+\epsilon)^{1/\Delta} }\right) d\theta
\right]\right)
\end{align}

We proceed the proof as follows:
\begin{enumerate}
\item Using the fact that $e^{-x}\geq \max\{0,1-x\}$, we obtain
\begin{align}\notag
\mathbf{Pr}\left(\hat{F}_{(\alpha),\min} \geq (1+\epsilon)F_{(\alpha)}\right)
\leq \exp\left(k \log \left[1-\frac{1}{\pi}\int_0^{\theta_0} 1- \frac{ g(\theta;\Delta) }{(1+\epsilon)^{1/\Delta} } d\theta,
\right]\right)
\end{align}
where $\theta_0$ is the solution to
\begin{align}\notag
1= \frac{ g(\theta;\Delta) }{(1+\epsilon)^{1/\Delta} }
\end{align}
\item We prove a more general result to solve for
\begin{align}\notag
\Delta^\gamma = \frac{ g(\theta_\gamma;\Delta) }{(1+\epsilon)^{1/\Delta} }.
\end{align}
We show the asymptotic expression for $\theta_\gamma$ is, as $\Delta\rightarrow0$,
\begin{align}\label{eqn_theta_gamma}
\theta_\gamma=& \pi - \pi \frac{\Delta}
{\Delta+\gamma\Delta\log\Delta+\log(1+\epsilon)+\Delta\log\left(\frac{1}{\gamma\Delta\log\Delta+\log(1+\epsilon)}+1\right)+O\left(\Delta^2\right)
}\\\notag
\end{align}
\item We approximate the integral $\int_0^{\theta_0} 1- \frac{ g(\theta;\Delta) }{(1+\epsilon)^{1/\Delta} } d\theta$ by the trapezoid rule.  Because $g(\theta,\Delta)$ is a convex function of $\theta$ as proved in Lemma \ref{lem_g}, we know this approximation still leads to an upper bound we are after.
\item To apply the trapezoid rule, it turns out that it suffices to use only one interior point, $\theta=\theta_1$, in addition to the two end points, $\theta=0 = \theta_{\infty}$ and $\theta=\theta_0$. $\theta_1$ is the solution to $\Delta = \frac{ g(\theta_1;\Delta) }{(1+\epsilon)^{1/\Delta} }$.
\item We can slightly improve the bound by using more points when applying the trapezoid rule, for example, $\theta = \theta_{1/2}$, in addition to $\theta_0$, $\theta_1$, and $\theta_\infty$.
\end{enumerate}

We defer the proof of (\ref{eqn_theta_gamma}) to Appendix \ref{app_proof_theta_gamma}. Assuming (\ref{eqn_theta_gamma}) holds, we have
\begin{align}\notag
&\mathbf{Pr}\left(\hat{F}_{(\alpha),\min} \geq (1+\epsilon)F_{(\alpha)}\right)\\\notag
=&\exp\left(k \log \left[1-\frac{1}{\pi}\int_0^\pi \exp\left(- \frac{ g(\theta;\Delta) }{(1+\epsilon)^{1/\Delta} }\right) d\theta
\right]\right)\\\notag
\leq&\exp\left(k \log \left[1-\frac{1}{\pi}\int_0^{\theta_0} 1- \frac{ g(\theta;\Delta) }{(1+\epsilon)^{1/\Delta}}  d\theta
\right]\right)\\\notag
\leq&\exp\left(k \log \left[1-\frac{1}{\pi}\left[\theta_1 - \frac{1}{2}\theta_1\Delta +\frac{1}{2}(1-\Delta)(\theta_0-\theta_1)\right]\right]\right)\\\notag
=&\exp\left(k \log \left[1-\frac{1}{2\pi}\left[\theta_0+\theta_1-\Delta\theta_0\right]\right]\right)
\end{align}

\begin{align}\notag
&2-\frac{1}{\pi}\left[\theta_0+\theta_1-\Delta\theta_0\right]\\\notag
=&\frac{1}{1+\log\Delta + \frac{1}{\Delta}\log(1+\epsilon)+ \log\left(\frac{1}{\Delta\log\Delta+\log(1+\epsilon)}+1\right)+O\left(\Delta\right)}\\\notag
+&\frac{1}{1+ \frac{1}{\Delta}\log(1+\epsilon)+ \log\left(\frac{1}{\log(1+\epsilon)}+1\right)+O\left(\Delta\right)}\\\notag
+&\Delta\frac{\frac{1}{\Delta}\log(1+\epsilon)+ \log\left(\frac{1}{\log(1+\epsilon)}+1\right)+O\left(\Delta\right)}
{1+\frac{1}{\Delta}\log(1+\epsilon)+ \log\left(\frac{1}{\log(1+\epsilon)}+1\right)+O\left(\Delta\right)}\\\notag
=&\frac{\Delta}{\Delta+\Delta\log\Delta + \log(1+\epsilon)+ \Delta\log\left(\frac{1}{\Delta\log\Delta+\log(1+\epsilon)}+1 \right)+O\left(\Delta^2\right)}\\\notag
+&\frac{\Delta}{\Delta+ \log(1+\epsilon)+ \Delta\log\left(\frac{1}{\log(1+\epsilon)}+1\right)+O\left(\Delta^2\right)}+\Delta+O\left(\Delta^2\right)\\\notag
=&\Delta+\frac{\Delta}{\log(1+\epsilon)}+\frac{\Delta}{\Delta\log\Delta+\log(1+\epsilon)}+O\left(\Delta^2\right)
\end{align}

Therefore, if we require
\begin{align}\notag
&\mathbf{Pr}\left(\hat{F}_{(\alpha),\min} \geq (1+\epsilon)F_{(\alpha)}\right)\\\notag
\leq&\exp\left(k \log\frac{1}{2}\left[
\Delta+\frac{\Delta}{\log(1+\epsilon)}+\frac{\Delta}{\Delta\log\Delta+\log(1+\epsilon)}+O\left(\Delta^2\right)
 \right]\right)\\\notag
 \leq& \delta,
\end{align}
we obtain our main result, the sample complexity bound,
\begin{align}\notag
k \geq \frac{\log\frac{1}{\delta}}{\log\frac{1}{\Delta} -\log\left(
\frac{1}{2}+\frac{1}{2\log(1+\epsilon)}+\frac{1}{2\Delta\log\Delta+2\log(1+\epsilon)}+O\left(\Delta\right)
\right)}.
\end{align}

It turns out, the term $\Delta\log\Delta$ can be almost removed, by using one additional interior point when applying the trapezoid rule.  Note that $|\Delta\log\Delta|$ is almost as small as $\Delta$, but we do not want to simply ignore this term. \\

Using two interior points, $\theta_1$ and $\theta_t$, where $0<t<1$, we obtain

\begin{align}\notag
&\mathbf{Pr}\left(\hat{F}_{(\alpha),\min} \geq (1+\epsilon)F_{(\alpha)}\right)\\\notag
=&\exp\left(k \log \left[1-\frac{1}{\pi}\int_0^\pi \exp\left(- \frac{  \left[\sin\left(\alpha \theta \right)\right]^{\alpha/\Delta} }{(1+\epsilon)^{1/\Delta} \left[\sin \theta \right]^{1/\Delta}} \sin\left( \theta\Delta\right)\right) d\theta
\right]\right)\\\notag
\leq&\exp\left(k \log \left[1-\frac{1}{\pi}\int_0^{\theta_0} 1- \frac{  \left[\sin\left(\alpha \theta \right)\right]^{\alpha/\Delta} }{(1+\epsilon)^{1/\Delta} \left[\sin \theta \right]^{1/\Delta}} \sin\left( \theta\Delta\right) d\theta
\right]\right)\\\notag
\leq&\exp\left(k \log \left[1-\frac{1}{\pi}\left[\theta_1 - \frac{1}{2}\theta_1\Delta +\frac{1}{2}\left(\theta_t-\theta_1\right)(1-\Delta+1-\Delta^t)+\frac{1}{2}(1-\Delta^t)(\theta_0-\theta_t)\right]\right]\right)\\\notag
=&\exp\left(k \log \left[1-\frac{1}{2\pi}\left[\theta_0+\theta_t-\Delta\theta_t-\Delta^t\theta_0+\Delta^t\theta_1\right]\right]\right)\\\notag
\end{align}

\begin{align}\notag
&2-\frac{1}{\pi}\left[\theta_0+\theta_t-\Delta\theta_t-\Delta^t\theta_0+\Delta^t\theta_1\right]\\\notag
=&\frac{1}{1+t\log\Delta + \frac{1}{\Delta}\log(1+\epsilon)+ \log\left(\frac{1}{t\Delta\log\Delta+\log(1+\epsilon)}+1\right)+O\left(\Delta\right)}\\\notag
+&\frac{1}{1+ \frac{1}{\Delta}\log(1+\epsilon)+ \log\left(\frac{1}{\log(1+\epsilon)}+1\right)+O\left(\Delta\right)}\\\notag
+&\Delta\frac{t\log\Delta+\frac{1}{\Delta}\log(1+\epsilon)+ \log\left(\frac{1}{t\Delta\log\Delta+\log(1+\epsilon)}+1\right)+O\left(\Delta\right)}
{1+t\log\Delta+\frac{1}{\Delta}\log(1+\epsilon)+ \log\left(\frac{1}{t\log\Delta+\log(1+\epsilon)}+1\right)+O\left(\Delta\right)}\\\notag
+&\Delta^t\frac{\frac{1}{\Delta}\log(1+\epsilon)+ \log\left(\frac{1}{\log(1+\epsilon)}+1\right)+O\left(\Delta\right)}
{1+\frac{1}{\Delta}\log(1+\epsilon)+ \log\left(\frac{1}{\log(1+\epsilon)}+1\right)+O\left(\Delta\right)}\\\notag
-&\Delta^t\frac{\log\Delta+\frac{1}{\Delta}\log(1+\epsilon)+ \log\left(\frac{1}{\Delta\log\Delta+\log(1+\epsilon)}+1\right)+O\left(\Delta\right)}
{1+\log\Delta+\frac{1}{\Delta}\log(1+\epsilon)+ \log\left(\frac{1}{\Delta\log\Delta+\log(1+\epsilon)}+1\right)+O\left(\Delta\right)}\\\notag
=&\Delta+\frac{\Delta}{\log(1+\epsilon)}+\frac{\Delta}{t\Delta\log\Delta+\log(1+\epsilon)}+O\left(\Delta^2\right)
\end{align}

Note that, if we choose $t$ to be too small (too close to 0), then $\left(-\Delta^t\theta_0+\Delta^t\theta_1\right)$ will be larger than $O\left(\Delta^2\right)$ and can not be ignored. Therefore, although we can minimize the impact of the term $\Delta\log\Delta$ to a very large extent, it can not be entirely removed, theoretically speaking.

\section{Conclusion}

Real-world data are often dynamic and can be modeled as data streams. Measuring  summary statistics of data streams such as the Shannon entropy has become an important task in many applications, for example, detecting anomaly events in large-scale networks. One line of active research is to approximate the Shannon entropy using the $\alpha$th frequency moments of the stream with $\alpha$ extremely close to 1.

Efficiently approximating the $\alpha$th frequency moments of data streams has been very heavily studied in theoretical computer science and databases. When $0<\alpha\leq2$, it is well-known that efficient $O\left(1/\epsilon^2\right)$-space algorithms exist, for example, {\em symmetric stable random projections}\cite{Article:Indyk_JACM06,Proc:Li_SODA08}, which however are impractical for estimating Shannon entropy using $\alpha$ extremely close to 1. Recently, \cite{Proc:Li_SODA09} provided an algorithm to achieve the $O\left(1/\epsilon\right)$  bound in the neighborhood of $\alpha=1$, based on the idea of {\em maximally-skewed stable random projections} (also called {\em Compressed Counting (CC)}). The $O\left(1/\epsilon\right)$ bound, although a very large improvement over the previous $O\left(1/\epsilon^2\right)$ bound, is still impractical.

This study proposes a new algorithm for CC based on the {\em sample minimum}, which is simple, practical, and still has very nice theoretical properties. Using this algorithm, we have proved that the sample complexity is essentially $O\left(\frac{1}{\log1/(1-\alpha) - \log1/\epsilon}\right)$ as $\alpha\rightarrow 1-$. This is a very large improvement over the previous $O(1/\epsilon)$ bound and may impact the practice.

\appendix

\section{Proof of Lemma \ref{lem_g}}\label{app_proof_lem_g}

For $\theta\in(0,\pi)$, let
\begin{align}\notag
g(\theta;\Delta) = \frac{  \left[\sin\left(\alpha \theta \right)\right]^{\alpha/\Delta} }{\left[\sin \theta \right]^{1/\Delta}} \sin\left( \theta\Delta\right).
\end{align}
It is easy to show that, as $\theta\rightarrow 0+$,
\begin{align}\notag
\lim_{\theta\rightarrow 0+} g(\theta,\Delta) =& \lim_{\theta\rightarrow 0+}\frac{  \left[\sin\left(\alpha \theta \right)\right]^{\alpha/\Delta} }{\left[\sin \theta \right]^{1/\Delta}} \sin\left( \theta\Delta\right)\\\notag
=&\lim_{\theta\rightarrow 0+} \left(\frac{\sin\left(\alpha \theta \right) }{\sin \theta} \right)^{1/\Delta} \frac{\sin\left( \theta\Delta\right)}{\sin\left(\alpha \theta \right)}\\\notag
=&\alpha^{1/\Delta}\frac{\Delta}{\alpha} = \Delta\alpha^{1/\Delta-1}.
\end{align}

The proof of the monotonicity of $g(\theta,\Delta)$ is omitted, because it is can be inferred from the proof of the convexity.

To show $g(\theta;\Delta)$ is a convex function $\theta$, it suffices to show it is log-convex. Since
\begin{align}\notag
g(\theta;\Delta) = \sin(\theta\Delta)\frac{ [\sin(\alpha\theta)]^{\alpha/\Delta}}{[\sin(\theta)]^{1/\Delta}}
=\frac{\sin(\theta\Delta)}{\sin(\alpha\theta)}\left[\frac{\sin(\alpha\theta)}{\sin(\theta)}\right]^{1/\Delta}
\end{align}
it suffices to show that both $\frac{\sin(\theta\Delta)}{\sin(\alpha\theta)}$ and $\left[\frac{\sin(\alpha\theta)}{\sin(\theta)}\right]^{1/\Delta}$ are log-convex.

\begin{align}\notag
&\frac{\partial \log \sin(\theta\Delta) - \log \sin(\alpha\theta)}{\partial \theta}
=\frac{\cos(\theta\Delta)}{\sin(\theta\Delta)}\Delta - \frac{\cos(\alpha\theta)}{\sin(\alpha\theta)}\alpha
\end{align}
\begin{align}\notag
&\frac{\partial^2 \log \sin(\theta\Delta) - \log \sin(\alpha\theta)}{\partial \theta^2}
=-\frac{\Delta^2}{\sin^2(\theta\Delta)} + \frac{\alpha^2}{\sin^2(\alpha\theta)}
=\left(\frac{\alpha}{\sin(\alpha\theta)}-\frac{\Delta}{\sin(\theta\Delta)}\right)
\left(\frac{\alpha}{\sin(\alpha\theta)}+\frac{\Delta}{\sin(\theta\Delta)}\right)
\end{align}

\begin{align}\notag
\frac{\partial \alpha\sin(\theta\Delta)-\Delta\sin(\alpha\theta)}{\partial \theta} = \Delta\alpha(\cos(\theta\Delta)-\cos(\alpha\theta))\geq 0 \hspace{0.5in} (\text{because} \ \ \Delta<0.5)
\end{align}
Therefore, $\alpha\sin(\theta\Delta)-\Delta\sin(\alpha\theta)\geq 0$ and $\frac{\sin(\theta\Delta)}{\sin(\alpha\theta)}$ is convex.

\begin{align}\notag
&\frac{\partial \log \sin(\alpha\theta) - \log \sin(\theta)}{\partial \theta}
=\frac{\cos(\alpha\theta)}{\sin(\alpha\theta)}\alpha - \frac{\cos(\theta)}{\sin(\theta)}
\end{align}
\begin{align}\notag
&\frac{\partial^2 \log \sin(\alpha\theta) - \log \sin(\theta)}{\partial \theta^2}
=-\frac{\alpha^2}{\sin^2(\alpha\theta)} + \frac{1}{\sin^2(\theta)}
=\left(\frac{1}{\sin(\theta)}-\frac{\alpha}{\sin(\alpha\theta)}\right)
\left(\frac{1}{\sin(\theta)}+\frac{\alpha}{\sin(\alpha\theta)}\right)
\end{align}

\begin{align}\notag
\frac{\partial \sin(\alpha\theta)-\alpha\sin(\theta)}{\partial \theta} = \alpha(\cos(\alpha\theta)-\cos(\theta))\geq 0 \hspace{0.5in} (\text{because} \ \alpha = 1-\Delta>0.5)
\end{align}
Therefore,  we have proved the convexity of $g\left(\theta;\Delta\right)$.

\section{Proof of  Equation (\ref{eqn_theta_gamma})}\label{app_proof_theta_gamma}

$\theta_\gamma$ is the solution to
\begin{align}\notag
 \Delta^\gamma = \frac{  \left[\sin\left(\alpha \theta \right)\right]^{\alpha/\Delta} }{(1+\epsilon)^{1/\Delta} \left[\sin \theta \right]^{1/\Delta}} \sin\left( \theta\Delta\right),
\end{align}
Equivalently,
\begin{align}\notag
\gamma\log\Delta + \frac{1}{\Delta}\log(1+\epsilon)+\frac{1}{\Delta}\log \sin\theta = \frac{1-\Delta}{\Delta}\log \sin\left(\theta - \Delta\theta\right) + \log \sin\left(\Delta\theta\right)
\end{align}
$\Longleftrightarrow$
\begin{align}\notag
\gamma\log\Delta + \frac{1}{\Delta}\log(1+\epsilon)+\log \frac{ \sin\left(\theta - \Delta\theta\right)}{ \sin\left(\Delta\theta\right)} = \frac{1}{\Delta}\log \frac{\sin\left(\theta - \Delta\theta\right)}{\sin\theta}
\end{align}
$\Longleftrightarrow$
\begin{align}\notag
\gamma\log\Delta + \frac{1}{\Delta}\log(1+\epsilon)+\log \left( \sin\theta\frac{\cos(\Delta\theta)}{\sin(\Delta\theta)} - \cos\theta\right) = \frac{1}{\Delta}\log
 \left(\cos(\Delta\theta)- \sin(\Delta\theta)\frac{\cos\theta}{\sin\theta} \right).
\end{align}
We apply Taylor expansions,
\begin{align}\notag
\gamma\log\Delta + \frac{1}{\Delta}\log(1+\epsilon)+ \log \left( \frac{-\sin\theta}{\Delta\theta\cos\theta} +\frac{\Delta\theta\sin\theta}{3\cos\theta} + 1+...\right)+\log\left(-\cos\theta\right) =-\frac{\Delta\theta^2}{2} - \frac{\theta\cos\theta}{\sin\theta} - \frac{\Delta\theta^2}{2}\frac{\cos^2\theta}{\sin^2\theta} +...
\end{align}
to obtain
\begin{align}\notag
\gamma\log\Delta + \frac{1}{\Delta}\log(1+\epsilon)+ \log \left( \frac{-\sin\theta}{\Delta\theta\cos\theta} +O\left(\Delta\right) + 1\right)+O\left(\Delta^2\right) =- \frac{\theta\cos\theta}{\sin\theta} +O\left(\Delta\right)
\end{align}
where we have replaced $\log\left(-\cos\theta\right)$ with $O\left(\Delta^2\right)$ (as $\Delta\rightarrow0$). This fact can be later verified.

Let $T = -\frac{\theta\cos\theta}{\sin\theta}$, $C = \gamma\log\Delta + \frac{1}{\Delta}\log(1+\epsilon)$. This requires us to solve a fixed point equation:
\begin{align}\notag
T = C + \log\left(\frac{1}{\Delta T} + O\left(\Delta\right)+1\right)+O\left(\Delta\right).
\end{align}

We resort to an iterative method.
Starting with $T^{(0)}=1$,
\begin{align}\notag
T^{(1)} = C + \log \left( \frac{1}{\Delta} +O\left(\Delta\right) + 1\right)+O\left(\Delta\right)=C-\log(\Delta)+O\left(\Delta\right).
\end{align}
\begin{align}\notag
T^{(2)} =& C + \log\left(\frac{1}{\Delta \left(C-\log(\Delta)+O\left(\Delta\right)\right)}+O\left(\Delta\right)+1\right)+O\left(\Delta\right)\\\notag
=&C + \log\left(\frac{1}{(\gamma-1)\Delta\log\Delta+\log(1+\epsilon) + O\left(\Delta^2\right)}+O\left(\Delta\right)+1\right)+O\left(\Delta\right)\\\notag
=&C +
\log\frac{1+(\gamma-1)\Delta\log\Delta+\log(1+\epsilon)+O(\Delta^2)}{
(\gamma-1)\Delta\log\Delta+\log(1+\epsilon)+O(\Delta^2)}
+O\left(\Delta\right)
\end{align}
\begin{align}\notag
T^{(3)} =& C + \log\left(\frac{1}{\Delta\left(
C +\log\frac{1+(\gamma-1)\Delta\log\Delta+\log(1+\epsilon)+O(\Delta^2)}{
(\gamma-1)\Delta\log\Delta+\log(1+\epsilon)+O(\Delta^2)}\right)}+O\left(\Delta\right)+1
\right)+O\left(\Delta\right)\\\notag
=&C +\log\left(\frac{1}{\gamma\Delta\log\Delta+\log(1+\epsilon)+O\left(\Delta\right)}+O\left(\Delta\right)+1\right)+O\left(\Delta\right)\\\notag
=&C +\log\left(\frac{1+\gamma\Delta\log\Delta+\log(1+\epsilon)+O\left(\Delta\right)}{\gamma\Delta\log\Delta+\log(1+\epsilon)+O\left(\Delta\right)}\right)+O\left(\Delta\right)
\end{align}
\begin{align}\notag
T^{(4)} =& C + \log\left(\frac{1}{\Delta\left(
C +\log\frac{1+\gamma\Delta\log\Delta+\log(1+\epsilon)+O(\Delta^2)}{
\gamma\Delta\log\Delta+\log(1+\epsilon)+O(\Delta^2)}\right)}+O\left(\Delta\right)+1
\right)+O\left(\Delta\right)\\\notag
=&C +\log\left(\frac{1}{\gamma\Delta\log\Delta+\log(1+\epsilon)+O\left(\Delta\right)}+O\left(\Delta\right)+1\right)+O\left(\Delta\right)\\\notag
=&C +\log\left(\frac{1+\gamma\Delta\log\Delta+\log(1+\epsilon)+O\left(\Delta\right)}{\gamma\Delta\log\Delta+\log(1+\epsilon)+O\left(\Delta\right)}\right)+O\left(\Delta\right).
\end{align}

At this point, we have reached an equilibrium. Therefore, we know
\begin{align}\notag
T =&\gamma\log\Delta+\frac{1}{\Delta}\log(1+\epsilon) +\log\left(\frac{1+\gamma\Delta\log\Delta+\log(1+\epsilon)}{\gamma\Delta\log\Delta+\log(1+\epsilon)}\right)+O\left(\Delta\right).
\end{align}
Note that
\begin{align}\notag
T =-\frac{\theta\cos\theta}{\sin\theta}=
\frac{\theta\cos\left(\pi-\theta\right)}{\sin(\pi-\theta)}=\theta\left(
\frac{1}{\pi-\theta} - \frac{\pi-\theta}{3} + O\left(\left(\pi-\theta\right)^3\right)\right) =
\frac{\theta}{\pi-\theta} + O\left(\pi-\theta\right)
\end{align}
Thus, assuming $O(\pi-\theta_r) = O\left(\Delta\right)$ (which can be verified), we obtain
\begin{align}\notag
\theta_\gamma =& \pi \frac{\gamma\log\Delta+\frac{1}{\Delta}\log(1+\epsilon)+\log\left(\frac{1}{\gamma\Delta\log\Delta+\log(1+\epsilon)}+1\right)+O\left(\Delta\right)
}
{1+\gamma\log\Delta+\frac{1}{\Delta}\log(1+\epsilon)+\log\left(\frac{1}{\gamma\Delta\log\Delta+\log(1+\epsilon)}+1\right)+O\left(\Delta\right)
}\\\notag
=& \pi \frac{\gamma\Delta\log\Delta+\log(1+\epsilon)+\Delta\log\left(\frac{1}{\gamma\Delta\log\Delta+\log(1+\epsilon)}+1\right)+O\left(\Delta^2\right)
}
{\Delta+\gamma\Delta\log\Delta+\log(1+\epsilon)+\Delta\log\left(\frac{1}{\gamma\Delta\log\Delta+\log(1+\epsilon)}+1\right)+O\left(\Delta^2\right)
}\\\notag
=& \pi - \pi \frac{\Delta}
{\Delta+\gamma\Delta\log\Delta+\log(1+\epsilon)+\Delta\log\left(\frac{1}{\gamma\Delta\log\Delta+\log(1+\epsilon)}+1\right)+O\left(\Delta^2\right)
}
\end{align}
To complete the proof, we must verify $O(\pi-\theta_r) = O\left(\Delta\right)$ and $\log(-\cos(\theta_r)) = O\left(\Delta^2\right)$. Indeed,
\begin{align}\notag
O(\pi-\theta_r) = \pi \frac{\Delta}
{\Delta+\gamma\Delta\log\Delta+\log(1+\epsilon)+\Delta\log\left(\frac{1}{\gamma\Delta\log\Delta+\log(1+\epsilon)}+1\right)+O\left(\Delta^2\right)
} = O\left(\Delta\right)
\end{align}
\begin{align}\notag
\log(-\cos(\theta_r)) = \log\left(\cos(\pi-\theta_r)\right) = \log\left(\cos(O(\Delta))\right) =\log\left(1-\frac{O(\Delta^2)}{2}\right) = O\left(\Delta^2\right).
\end{align}

%


\begin{thebibliography}{10}

\bibitem{Proc:Aggarwal_KDD04}
Charu~C. Aggarwal, Jiawei Han, Jianyong Wang, and Philip~S. Yu.
\newblock On demand classification of data streams.
\newblock In {\em KDD}, pages 503--508, Seattle, WA, 2004.

\bibitem{Proc:Alon_STOC96}
Noga Alon, Yossi Matias, and Mario Szegedy.
\newblock The space complexity of approximating the frequency moments.
\newblock In {\em STOC}, pages 20--29, Philadelphia, PA, 1996.

\bibitem{Article:Chakrabarti_06}
Khanh Do~Ba Amit~Chakrabarti and S.~Muthukrishnan.
\newblock Estimating entropy and entropy norm on data streams.
\newblock {\em Internet Mathematics}, 3(1):63--78, 2006.

\bibitem{Proc:Babcock_PODS02}
Brian Babcock, Shivnath Babu, Mayur Datar, Rajeev Motwani, and Jennifer Widom.
\newblock Models and issues in data stream systems.
\newblock In {\em PODS}, pages 1--16, Madison, WI, 2002.

\bibitem{Proc:Kumar_FOCS02}
Ziv Bar-Yossef, T.~S. Jayram, Ravi Kumar, and D.~Sivakumar.
\newblock An information statistics approach to data stream and communication
  complexity.
\newblock In {\em FOCS}, pages 209--218, Vancouver, BC, Canada, 2002.

\bibitem{Proc:Bhuvanagiri_ESA06}
Lakshminath Bhuvanagiri and Sumit Ganguly.
\newblock Estimating entropy over data streams.
\newblock In {\em ESA}, pages 148--159, 2006.

\bibitem{Proc:Brauckhoff_IMC06}
Daniela Brauckhoff, Bernhard Tellenbach, Arno Wagner, Martin May, and Anukool
  Lakhina.
\newblock Impact of packet sampling on anomaly detection metrics.
\newblock In {\em IMC}, pages 159--164, 2006.

\bibitem{Proc:Chakrabarti_SODA07}
Amit Chakrabarti, Graham Cormode, and Andrew McGregor.
\newblock A near-optimal algorithm for computing the entropy of a stream.
\newblock In {\em SODA}, pages 328--335, 2007.

\bibitem{Article:Chambers_JASA76}
John~M. Chambers, C.~L. Mallows, and B.~W. Stuck.
\newblock A method for simulating stable random variables.
\newblock {\em Journal of the American Statistical Association},
  71(354):340--344, 1976.

\bibitem{Proc:Domeniconi_ICDM01}
Carlotta Domeniconi and Dimitrios Gunopulos.
\newblock Incremental support vector machine construction.
\newblock In {\em ICDM}, pages 589--592, San Jose, CA, 2001.

\bibitem{Proc:Feigenbaum_FOCS99}
Joan Feigenbaum, Sampath Kannan, Martin Strauss, and Mahesh Viswanathan.
\newblock An approximate $l_1$-difference algorithm for massive data streams.
\newblock In {\em FOCS}, pages 501--511, New York, 1999.

\bibitem{Proc:Feinstein_DARPA03}
Laura Feinstein, Dan Schnackenberg, Ravindra Balupari, and Darrell Kindred.
\newblock Statistical approaches to \text{DDoS} attack detection and response.
\newblock In {\em DARPA Information Survivability Conference and Exposition},
  pages 303--314, 2003.

\bibitem{Article:Flajolet_BIT85}
Philippe Flajolet.
\newblock Approximate counting: A detailed analysis.
\newblock {\em BIT}, 25(1):113--134, 1985.

\bibitem{Proc:Ganguly_RANDOM07}
Sumit Ganguly and Graham Cormode.
\newblock On estimating frequency moments of data streams.
\newblock In {\em APPROX-RANDOM}, pages 479--493, Princeton, NJ, 2007.

\bibitem{Proc:Guha_SODA06}
Sudipto Guha, Andrew McGregor, and Suresh Venkatasubramanian.
\newblock Streaming and sublinear approximation of entropy and information
  distances.
\newblock In {\em SODA}, pages 733 -- 742, Miami, FL, 2006.

\bibitem{Proc:Harvey_FOCS08}
Nicholas J.~A. Harvey, Jelani Nelson, and Krzysztof Onak.
\newblock Sketching and streaming entropy via approximation theory.
\newblock In {\em FOCS}, 2008.

\bibitem{Proc:Harvey_ITW08}
Nicholas J.~A. Harvey, Jelani Nelson, and Krzysztof Onak.
\newblock Streaming algorithms for estimating entropy.
\newblock In {\em ITW}, 2008.

\bibitem{Article:Havrda_67}
M~E. Havrda and F.~Charv\'at.
\newblock Quantification methods of classification processes: Concept of
  structural $\alpha$-entropy.
\newblock {\em Kybernetika}, 3:30--35, 1967.

\bibitem{Book:Henzinger_99}
Monika~R. Henzinger, Prabhakar Raghavan, and Sridhar Rajagopalan.
\newblock {\em Computing on Data Streams}.
\newblock American Mathematical Society, Boston, MA, USA, 1999.

\bibitem{Article:Indyk_JACM06}
Piotr Indyk.
\newblock Stable distributions, pseudorandom generators, embeddings, and data
  stream computation.
\newblock {\em Journal of ACM}, 53(3):307--323, 2006.

\bibitem{Proc:Indyk_STOC05}
Piotr Indyk and David~P. Woodruff.
\newblock Optimal approximations of the frequency moments of data streams.
\newblock In {\em STOC}, pages 202--208, Baltimore, MD, 2005.

\bibitem{Proc:Lakhina_SIGCOMM05}
Anukool Lakhina, Mark Crovella, and Christophe Diot.
\newblock Mining anomalies using traffic feature distributions.
\newblock In {\em SIGCOMM}, pages 217--228, Philadelphia, PA, 2005.

\bibitem{Proc:Lall_SIGMETRICS06}
Ashwin Lall, Vyas Sekar, Mitsunori Ogihara, Jun Xu, and Hui Zhang.
\newblock Data streaming algorithms for estimating entropy of network traffic.
\newblock In {\em SIGMETRICS}, pages 145--156, 2006.

\bibitem{Proc:Li_SODA08}
Ping Li.
\newblock Estimators and tail bounds for dimension reduction in $l_\alpha$
  ($0<\alpha\leq 2$) using stable random projections.
\newblock In {\em SODA}, pages 10 -- 19, San Francisco, CA, 2008.

\bibitem{Proc:Li_SODA09}
Ping Li.
\newblock Compressed counting.
\newblock In {\em SODA}, New York, NY, 2009.

\bibitem{Proc:Li_UAI09}
Ping Li.
\newblock Improving compressed counting.
\newblock In {\em UAI}, Montreal, CA, 2009.

\bibitem{Proc:Mei_WSDM08}
Qiaozhu Mei and Kenneth Church.
\newblock Entropy of search logs: How hard is search? with personalization?
  with backoff?
\newblock In {\em WSDM}, pages 45 -- 54, Palo Alto, CA, 2008.

\bibitem{Article:Morris_CACM78}
Robert Morris.
\newblock Counting large numbers of events in small registers.
\newblock {\em Commun. ACM}, 21(10):840--842, 1978.

\bibitem{Article:Muthukrishnan_05}
S.~Muthukrishnan.
\newblock Data streams: Algorithms and applications.
\newblock {\em Foundations and Trends in Theoretical Computer Science},
  1:117--236, 2 2005.

\bibitem{Article:Paninski_NC03}
Liam Paninski.
\newblock Estimation of entropy and mutual information.
\newblock {\em Neural Comput.}, 15(6):1191--1253, 2003.

\bibitem{Proc:Renyi_61}
Alfred R\'enyi.
\newblock On measures of information and entropy.
\newblock In {\em The 4th Berkeley Symposium on Mathematics, Statistics and
  Probability 1960}, pages 547--561, 1961.

\bibitem{Proc:Saks_STOC02}
Michael~E. Saks and Xiaodong Sun.
\newblock Space lower bounds for distance approximation in the data stream
  model.
\newblock In {\em STOC}, pages 360--369, Montreal, Quebec, Canada, 2002.

\bibitem{Article:Tsallis_88}
Constantino Tsallis.
\newblock Possible generalization of boltzmann-gibbs statistics.
\newblock {\em Journal of Statistical Physics}, 52:479--487, 1988.

\bibitem{Proc:Woodruff_SODA04}
David~P. Woodruff.
\newblock Optimal space lower bounds for all frequency moments.
\newblock In {\em SODA}, pages 167--175, New Orleans, LA, 2004.

\bibitem{Proc:Xu_SIGCOMM05}
Kuai Xu, Zhi-Li Zhang, and Supratik Bhattacharyya.
\newblock Profiling internet backbone traffic: behavior models and
  applications.
\newblock In {\em SIGCOMM '05: Proceedings of the 2005 conference on
  Applications, technologies, architectures, and protocols for computer
  communications}, pages 169--180, 2005.

\bibitem{Article:ICDM10}
Qiang Yang and Xingdong Wu.
\newblock 10 challeng problems in data mining research.
\newblock {\em International Journal of Information Technology and Decision
  Making}, 5(4):597--604, 2006.

\bibitem{Proc:Zhao_IMC07}
Haiquan Zhao, Ashwin Lall, Mitsunori Ogihara, Oliver Spatscheck, Jia Wang, and
  Jun Xu.
\newblock A data streaming algorithm for estimating entropies of od flows.
\newblock In {\em IMC}, San Diego, CA, 2007.

\end{thebibliography}

\end{document}